\documentclass[aps,preprint,aps,showpacs]{revtex4}

\begin{document}

\title{Accurate first principles detailed balance determination of Auger
recombination and impact ionization rates in semiconductors}
\author{S. Picozzi}
\affiliation{Istituto Nazionale di Fisica della Materia (INFM), Dip. Fisica, Universit\`a
degli Studi di L'Aquila, 67010 Coppito (L'Aquila), Italy }
\author{R.Asahi}
\affiliation{Toyota Central R $\&$ D Labs., Inc., Nagakute, Japan}
\author{C.B.Geller}
\affiliation{Bettis Atomic Power Laboratory, West Mifflin, PA 15122, USA}
\author{A. J. Freeman}
\affiliation{Department of Physics and Astronomy and Materials Research Center,
Northwestern University, Evanston, IL 60208 (U.S.A.)}

\begin{abstract}
The technologically important problem of predicting Auger recombination
lifetimes in semiconductors is addressed by means of a fully
first--principles formalism. The calculations employ highly precise energy
bands and wave functions provided by the full--potential linearized
augmented plane wave (FLAPW) code based on the screened exchange local
density approximation. The minority carrier Auger lifetime is determined by
two closely related approaches: \emph{i}) a direct evaluation of the Auger
rates within Fermi's Golden Rule, and \emph{ii}) an indirect evaluation,
based on a detailed balance formulation combining Auger recombination and
its inverse process, impact ionization, in a unified framework. Calculated
carrier lifetimes determined with the direct and indirect methods show
excellent consistency \emph{i}) between them for $n$-doped GaAs and \emph{ii}%
) with measured values for GaAs and InGaAs. This demonstrates the validity
and accuracy of the computational formalism for the Auger lifetime and
indicates a new sensitive tool for possible use in materials performance
optimization. 
\end{abstract}

\maketitle


Led by advances in computational simulation, the paradigm within
semiconductor device engineering is shifting away from the static
perspective of material properties as design \textquotedblleft constraints"
towards a more empowering view of material properties as design optimization
parameters. This growing trend is evident clearly in recent initiatives in
nano--engineered materials from quantum dots, quantum wells and nanowires to
band structure engineering of conventional \cite{apl} and ordered
superlattice \cite{azaf} materials. However, despite the fundamental
importance of these processes, owing to a lack of adequate capability for
their prediction, carrier recombination processes in semiconductors have
remained largely beyond the reach of materials design. Minority carrier
lifetime is a critical, performance-limiting material parameter in many
opto--electronic devices, as well as in bipolar transistors, $p-n$
rectifiers and thyristors. Two recombination processes - radiative and Auger
- generally impose theoretical and practical limits on carrier lifetimes.
Minority carrier lifetimes in heavily doped and/or narrow band gap materials
(such as 1.6 $\mu $m In$_{0.53}$Ga$_{0.47}$As - a basic material for
thermophotovoltaic devices) tend to be limited by Auger recombination (AR),
and so its first--principles determination is the focus of this paper.

In $n$-doped, direct band gap materials, the dominant Auger process is
electron--electron--hole ($e^{-}e^{-}h$) recombination\cite{landsberg}. As
shown in Fig. \ref{process} (a), in the $e^{-}e^{-}h$ process, a valence
band hole decays via non-radiative recombination with a conduction band
electron, conserving energy and momentum through transfer to a second
conduction band electron. This process commonly is referred to as the
conduction--conduction--conduction--``heavy hole'' (CCCH) process. In the
inverse process of electron-initiated impact ionization (I-I), depicted in
Fig. \ref{process} (b), an energetic primary electron promotes a secondary
valence band electron into the conduction band, producing a mobile
electron--hole pair. Processes closely analogous to those depicted in Fig. %
\ref{process} (a) and (b) occur in $p$-type materials, resulting in an
electron decay through a ``conduction electron"--``heavy hole''--``heavy
hole''--``light hole'' (CHHL) AR and in a hole--initiated I-I, respectively.

In this work, 
the fully first--principles approach that was used successfully to evaluate
I-I rates \cite{slv} is extended to the calculation of AR rates which are
found to be in good agreement with previous experimental and theoretical
values reported for the most studied semiconductor -- $n$--type GaAs -- and
for technologically important InGaAs. Moreover, AR lifetimes calculated
directly via Fermi's Golden Rule are shown to be strikingly consistent with
lifetime values determined via detailed balance principles proceeding from
calculated I-I rates. To our knowledge, this density functional theory (DFT%
\cite{dft}) study represents the first fully ab initio determination of
Auger recombination lifetimes. The agreement of computational results with
experiment for $n$--type GaAs and In$_{0.5}$Ga$_{0.5}$As indicates that this
method is sufficiently precise and reliable to be able to correctly predict
trends of the Auger coefficient as a function of physical parameters, such
as doping, composition, etc. Since the current approach is all-electron and
ab initio in nature, we expect the same methods to be equally successful for
new and more complex materials. Finally, we find that detailed balance
evaluations proceeding from I-I rate calculations appear to be a
computationally efficient means of determining AR rates.

Within an independent particle scheme\cite{laks}, given a hole in state $%
(n_2,\mathbf{k_2})$, the AR rate is expressed as:

\begin{center}
$R^{AR}(n_2,\mathbf{k_2}) = 2 \frac{2\pi}{\hbar} \sum_{n_3,n_4} \int d^3 
\mathbf{k_3} \int d^3 \mathbf{k_4} |M|^2 f(E^{n_4}_{\mathbf{k_4}})$
\end{center}

\begin{equation}
f(E^{n_3}_{\mathbf{k_3}})g(E^{n_1}_{\mathbf{k_1}}) \delta(E^{n_3}_{\mathbf{%
k_3}}+E^{n_4}_{\mathbf{k_4}}- E^{n_1}_{\mathbf{k_1}}- E^{n_2}_{\mathbf{k_4}})
\label{direct}
\end{equation}
and the total AR rate is: 
\begin{equation}
R^{AR} = \sum_{n_2} \int d^3 \mathbf{k_2}\:g(E^{n_2}_{\mathbf{k_2}})
\:R^{AR}(n_2,\mathbf{k_2})  \label{total}
\end{equation}

Here, the $n_{i}$ are the band indices ($n_{i},i$ = 1,...,7 ($i$ = 5,...,11)
for GaAs (InGaAs), corresponding to the four valence and three conduction
bands considered) and \textbf{k$_{1,3,4}$} are \textbf{k} points in the full
Brillouin zone (BZ) \cite{compdet}. The many--fold integration over the BZ
is performed according to the Sano--Yoshii scheme\cite{sy}, using 152 \cite%
{accuracy} and 126 \textbf{k} points in the irreducible wedge of the
zincblende and tetragonal Brillouin zone, respectively \cite{cuau}. The
summation over \textbf{k$_{2}$} (Eq.\ref{total}) is carried out in the
irreducible wedge of the zone. Energy conservation is enforced through the $%
\delta $ function in the energy eigenvalues; $f$ denotes the Fermi--Dirac
occupation probability and $g=1-f$. The matrix elements, taking into account
both \emph{direct} and \emph{exchange} terms (obtained from the direct
contribution through the exchange of final states), are of the form: $%
|M|^{2}=\frac{1}{2}(|M_{D}|^{2}+|M_{E}|^{2}+|M_{D}-M_{E}|^{2})$. The direct
term is given as:

\begin{eqnarray}
M_{D} &=&\frac{4\pi e^{2}}{\Omega }\sum_{\mathbf{G_{0},G_{U}}}\delta (%
\mathbf{k_{3}+k_{4}-k_{1}-k_{2}+G_{0}}) \\
&&\frac{\rho _{n_{3},\mathbf{k_{3}};n_{1},\mathbf{k_{1}}}(\mathbf{G_{U}}%
)\rho _{n_{4},\mathbf{k_{4}};n_{2},\mathbf{k_{2}}}(\mathbf{G_{0}-G_{U}})}{%
\varepsilon (q)(|\mathbf{k_{1}-k_{3}+G_{U}}|^{2}+\lambda ^{2})}  \nonumber
\end{eqnarray}

\noindent where $e$ is the electronic charge and $\Omega $ is the volume of
the unit cell. Momentum conservation is enforced through the $\delta $
function in the \textbf{k} vectors and $\rho _{n_{f},\mathbf{k_{f}};n_{i},%
\mathbf{k_{i}}}(\mathbf{G}) $ is the Fourier transform of the overlap matrix
of the wave functions, \emph{i.e.}, $\rho_{n_{f},\mathbf{k_{f}};n_{i},%
\mathbf{k_{i}}}(\mathbf{r})= \Psi _{n_{f},\mathbf{k_{f}}}^{*}(\mathbf{r}%
)\Psi _{n_{i},\mathbf{k_{i}}}(\mathbf{r})$. The subscripts $i$ and $f$
denote initial and final states; $q=|\mathbf{k_{1}-k_{3}+G_{U}}|$ is the
momentum transfer and $\mathbf{G_{0}}$ and $\mathbf{G_{U}}$ are reciprocal
lattice vectors. To describe the interaction between the valence and
conduction band electrons, we used a Coulomb interaction screened through a
static dielectric function proposed by Cappellini \emph{et al.}, $%
\varepsilon (q)$ \cite{cappe}, that has been shown to be very accurate for
semiconductors. The interaction between conduction band carriers is modeled
through a Debye potential (screened Coulomb interaction), in which the
inverse of the screening length is expressed as $\lambda =\sqrt{\frac{4\pi
n^{0}e^{2}}{K_{B}T}}$, $n^{0}$, $T$ and $K_{B}$ being the carrier density,
the temperature (here, T=300 K) and the Boltzmann constant, respectively.

The theoretical estimate of the I-I rate\cite{schattke,hab} is known to be
quite sensitive to the assumed band structure. Therefore, the importance of
employing accurate quasi-particle bands as a basis for I-I and AR
calculations needs to be emphasized. To faithfully reproduce experimental
band structures (both occupied and virtual states), we employ the
screened--exchange local density approximation (sX-LDA)\cite{kleinman,seidl}%
, as implemented\cite{ryoji} self--consistently in the highly precise,
all--electron, full--potential linearized augmented plane wave (FLAPW)\cite%
{FLAPW} method. The sX-LDA approach corrects known deficiencies of the LDA
for excited states and has been shown to be very successful for treating a
large variety of semiconductors\cite{ryoji,apl}.

The effect of spin--orbit coupling, neglected here, was investigated in Ref.%
\cite{hab} for GaAs; the overall effect was shown to be relatively small.
Umklapp processes for \textbf{G$_{U}$} = 0 were fully included in our
calculations; the \textbf{G$_{0}$} $\neq $ 0 terms, which are important at
high energies were not included, since Auger is a threshold process. 
Phonon--assisted transitions\cite{take_ph,dutta} and many-body\cite%
{take_pl,bude} effects were also not included; for InGaAs, Dutta and Nelson 
\cite{dutta} showed that phonon assisted AR should not be relevant compared
to the direct AR. Within these approximations, giving an estimated overall
error of 30-40 $\%$ on the AR rates, we show below that our simple
band-to--band AR within DFT correctly describes the physical trends as a
function of several technological parameters, and so may be used as a
sensitive tool in materials device optimization. 

It can be shown \cite{srini}, for band gaps 
$E_{gap} >> K_B T$, that the Auger partial lifetime can be expressed as

\begin{equation}
\frac{1}{\tau _{p}}=\frac{R^{AR}}{N_{p}^{0}}  \label{tau}
\end{equation}

\noindent where $N_{p}^{0}$ is the number of minority carriers per unit cell
at equilibrium ($N_{p}^{0}/\Omega $ corresponds to the minority carrier
density $p$)\cite{srini}. In Fig.\ref{lifetime} we show the dependence of
the calculated hole Auger lifetime on carrier density, 
which is approximately linear on a log--log scale. Similar trends have been
reported, for example, in an empirical pseudopotential study of AR in Si 
\cite{laks} and in a recent experimental study of AR in $n$-doped InGaAs 
\cite{ahrenkiel}.

The Auger coefficient, $C_{n}$, is defined by the relation: 
\begin{equation}
\frac{1}{\tau _{p}}=C_{n}(n^{0})^{2}  \label{cn}
\end{equation}

\smallskip

\noindent Recall that partial recombination lifetimes usually are
\textquotedblleft determined\textquotedblright\ experimentally based on
curve fits to carrier lifetime measurements resulting from several competing
recombination processes. The parabolic relationship above is the usual basis
for determinations of the Auger recombination coefficient. It is noted,
however, \cite{laks} that both shifts in the Fermi energy (particularly
pronounced in many III-V alloys with low effective masses), and changes in
the Debye screening length introduce a carrier density dependence into Eqs.%
\ref{direct} and \ref{total}, so that $C_{n}$ itself is expected to depend
on $n$. This is actually the case from our calculations. Indeed, in GaAs our 
$C_{n}$, determined as the ratio of the inverse lifetime and the carrier
density squared, 
ranges from 2.3x10$^{-30}$ cm$^{-6}$s$^{-1}$ at $n\sim $ 
4x10$^{15}$ cm$^{-3}$ to 0.5x10$^{-30}$ cm$^{-6}$s$^{-1}$ at $n\sim $ 
6x10$^{16}$ cm$^{-3}$. This range of values is compared in Table \ref{compar}
with previously reported experimental\cite{lush,strauss,marin} and
theoretical\cite{take,haug} results. In InGaAs, there is only a slight
dependence of $C_{n}$ on $n$, $C_{n}$ ranging from 0.8x10$^{-29}$ cm$^{-6}$s$%
^{-1}$ to 1.1x10$^{-29}$ cm$^{-6}$s$^{-1}$ for concentrations ranging from
7x10$^{16}$ cm$^{-3}$ to 2x10$^{18}$ cm$^{-3}$. These values are compared in
Table \ref{ingaas} with previous experimental \cite{henry,ahr_old,ahrenkiel}
and theoretical \cite{take_tie,bardy} results. The comparison between the
calculated values for the binary and the ternary compound shows that there
is a strong reduction (by an order of magnitude) of $C_{n}$, suggesting that
the Auger process is crucial in the narrow--gap InGaAs. Despite the wide
range over which the reported Auger coefficients vary, the calculated trend
with Ga content is confirmed experimentally, unambiguously emphasizing the
power of DFT in predicting both quantitatively and qualitatively correct
physical trends as a function of doping, pressure, composition, etc. The
wide range of reported Auger coefficients suggests the complexity of both
the measurements and the calculations. 
Our result for the Auger coefficient is in good agreement with most of the
published experimental values and is again consistent, when compared, with
previously reported results from model calculations incorporating various
simplifications (such as 
constant effective masses \cite{haug,take}). All considered, we conclude
that our results confirm the reliability of the computational methods and
band structures herein described \emph{i}) to accurately calculate Auger
recombination lifetimes that cannot be easily or unambiguously determined by
experiment; \emph{ii}) for mapping trends with respect to material
composition and dopant density, that are useful to guide optoelectronic
device and materials design; and \emph{iii}) for describing Auger
recombination in novel and more complex materials.

Auger recombination and impact ionization are related as inverse microscopic
processes through the principle of detailed balance (PDB), in the same
manner as optical emission and absorption. In fact, in the radiative
(non-radiative) case, the $e-h$ generation process is the absorption (I-I),
in which the initial object is the photon (highly energetic impacting
electron); correspondingly, their inverse recombination process is emission
(AR), in which the final product is the photon (Auger electron).

Following Landsberg\cite{landsberg}, 
the AR lifetime is given by,




\begin{equation}
\frac{1}{\tau _{p}}\sim \frac{1}{N_{p}^{0}}\sum_{n_{1}}\int d^{3}\mathbf{%
k_{1}}\beta (n_{1},\mathbf{k_{1}})R^{I-I}(n_{1},\mathbf{k_{1}})f(n_{1},%
\mathbf{k_{1}})  \label{vrs_slv}
\end{equation}

\noindent where $\beta (n_{1},\mathbf{k_{1}})$ is a factor ($\leq 1$) that
takes into account the occurrence of processes from state ($n_1,\mathbf{k_{1}%
}$) other than I-I and the $R^{I-I}(n_{1},\mathbf{k_{1}})$ are the I-I
rates. 
This last expression is the non-radiative equivalent of the van
Roosbroeck--Shockley relation \cite{vrs}, that has been used extensively in
determinations of radiative recombination lifetimes from the complex
dielectric function. However, while experimental dielectric function data
(or equivalently, refractive index and extinction coefficient data) are
abundant and highly reliable, experimental data for electron--initiated
impact ionization rates are normally inadequate to estimate the AR lifetime
from Eq. (\ref{vrs_slv}). A utility of the current formalism is that one can
calculate both I-I rates and AR rates either independently from first
principles, or from one another via detailed balance arguments. In our
formalism, we focus on two selected processes (I-I and AR), each being the
inverse of the other; therefore, the PDB holds with $\beta (n_1,\mathbf{k_1})
$ = 1 (or, equivalently, that every energetic electron in state $(n_1,%
\mathbf{k_1})$ initiates an impact ionization process). In Fig. \ref%
{lifetime}, Eq. (\ref{vrs_slv}) is applied to the previously calculated I-I
rates (see \cite{slv} for details) and the Auger lifetime for $n^0$ = 1.0x10$%
^{16}$ cm$^{-3}$ is derived. The open symbol in Fig. \ref{lifetime} denotes
the value obtained, which is in striking agreement with the values
calculated using the direct approach (see Eq.\ref{direct}). The excellent
agreement achieved between the direct and ''indirect'' calculations of both
parameters provides confidence in the equivalency of the two approaches and
in the numerical procedures embodied therein. 


In summary, we have presented a fully first--principles formalism for
calculating Auger recombination lifetimes using screened exchange FLAPW
quasiparticle wavefunctions and band structures determined
self-consistently. The numerical accuracy of the approach was checked using
the principle of detailed balance applied to the two simulated processes,
I-I and AR. The Auger lifetime was determined using two equivalent
approaches, \textquotedblleft direct\textquotedblright\ and
\textquotedblleft indirect,\textquotedblright\ with highly consistent
results. Our results for GaAs and InGaAs are in excellent agreement with the
most accurate experimental and theoretical data, providing confidence in the
computational method and justifying its future applications to more complex
systems.

Work at Northwestern University supported by the National Science Foundation
(through its MRSEC program at the Materials Research Center). 

\begin{table}[tbp]
\caption{Auger coefficients in $n$--type GaAs. The sX-LDA FLAPW results are
given to cover the range of possible $C_n$ values.}
\label{compar}%
\begin{tabular}{|c|c|}
\hline\hline
Exp.technique or theor.approach & $C_n$ (cm$^6$s$^{-1}$) \\ \hline\hline
sX-LDA FLAPW & (0.5 $\leq C_n \leq$ 2.3)$\cdot$10$^{-30}$ \\ \hline\hline
Photoacoustic determination$^a$ & 1.3$\cdot$10$^{-30}$ \\ 
Time--resolved photoluminescence decay$^b$ & (7$\pm$4)$\cdot$10$^{-30}$ \\ 
Time--resolved photoluminescence decay$^c$ & upper limit of 1.6$\cdot$10$%
^{-29}$ \\ \hline\hline
non-parab. bands + phonon-assisted + eff. masses$^d$ & 4.7$\cdot$10$^{-30}$
\\ 
non-parab. bands + phonon-assisted + $\mathbf{k\cdot p}$$^e$ & 1.5$\cdot$10$%
^{-31}$ \\ \hline\hline
\end{tabular}%
\par
\bigskip
\par
\noindent $a$. Ref. \onlinecite{marin}
\par
\noindent$b$. Ref. \onlinecite{strauss}
\par
\noindent$c$. Ref. \onlinecite{lush}
\par
\par
\noindent $d$. Ref. \onlinecite{haug}
\par
\noindent $e$. Ref. \onlinecite{take}
\end{table}

\begin{table}[tbp]
\caption{Auger coefficients in $n$--type InGaAs.}
\label{ingaas}%
\begin{tabular}{|c|c|}
\hline\hline
Exp.technique or theor.approach & $C_n$ (cm$^6$s$^{-1}$) \\ \hline\hline
sX-LDA FLAPW & (0.8 $\leq C_n \leq$ 1.1)$\cdot$10$^{-29}$ \\ \hline\hline
Time--resolved photoluminescence decay$^a$ & 0.5$\cdot$10$^{-29}$ \\ 
Radio-frequency photoconductive decay$^b$ & 7$\cdot$10$^{-29}$ \\ 
Photoluminescence photon--counting $^c$ & 18$\cdot$10$^{-29}$ \\ 
+ degenerate carrier conditions &  \\ \hline\hline
non-parab. bands + phonon-assisted + eff. masses$^d$ & 0.5$\cdot$10$^{-29}$
\\ 
Kane model for wave functions$^e$ & 3$\cdot$10$^{-29}$ \\ \hline\hline
\end{tabular}%
\par
\bigskip
\par
\noindent $a$. Ref. \onlinecite{henry}
\par
\noindent$b$. Ref. \onlinecite{ahr_old}
\par
\noindent$c$. Ref. \onlinecite{ahrenkiel}
\par
\noindent $d$. Ref. \onlinecite{take_tie}
\par
\noindent $e$. Ref. \onlinecite{bardy}
\par
\end{table}
\newpage

{Simulated processes in $n$--type materials: (a)
Auger recombination, resulting in a hole--decay; (b) electron initiated
impact ionization, resulting in a pair production.}

{Calculated hole Auger lifetime vs carrier
concentration in (a) GaAs and (b) InGaAs. The results for the binary
compound are obtained according to the direct approach (see Eqs.(2-5) -
filled circles) and to the indirect approach (see Eq.(6) - open circle).}



\begin{thebibliography}{99}
\bibitem{compdet} With these computational details, sufficient to obtain
accurate results - see Ref.\cite{slv} -, the calculations were performed on
a single 512 Mb RAM Dec Alpha workstation in approximately 300 CPU hours.
\bibitem{cuau} In$_{0.5}$Ga$_{0.5}$As was simulated using a [001]-ordered
1x1 InAs/GaAs superlattice, with four atoms in the unit cell.
\end{thebibliography}
\end{document}